\documentclass[conference]{IEEEtran}
\usepackage{amsmath}
\usepackage{amsthm}
\usepackage{amssymb}
\usepackage{bm}
\usepackage{xspace}
\usepackage{xcolor}
\usepackage{graphicx}
\usepackage{url}
\usepackage{cite}
\usepackage{framed}
\usepackage{float}
\usepackage{rotating}
\usepackage{subfigure}

\newcommand{\executeiffilenewer}[3]{%
\ifnum\pdfstrcmp{\pdffilemoddate{#1}}%
{\pdffilemoddate{#2}}>0%
{\immediate\write18{#3}}\fi%
}

\graphicspath{{images/}}

\newcommand{\RM}[1]{\MakeUppercase{\romannumeral #1{}}}

\newcounter{algocount}

\theoremstyle{plain}

\theoremstyle{definition}

\theoremstyle{plain}

\theoremstyle{definition}

\theoremstyle{remark}

\newcommand{\vecone}{\boldsymbol{1}}

\title{Strategies for Distributed Sensor Selection \\ Using Convex Optimization}

\IEEEoverridecommandlockouts

\author{\IEEEauthorblockN{Fabian Altenbach, Steven Corroy, Georg B\"ocherer\IEEEauthorrefmark{1}, and Rudolf Mathar}
\IEEEauthorblockA{Institute for Theoretical Information
Technology,
RWTH Aachen University, Germany\\
\IEEEauthorrefmark{1}Institute for Communications Engineering, Technische Universit\"at M\"unchen, Germany
\\ Email: \texttt{\{altenbach,corroy,mathar\}@ti.rwth-aachen.de}, \texttt{georg.boecherer@tum.de}}
\thanks{This work has been supported by the UMIC Research Center, RWTH
Aachen University.}
}

\begin{document}
\maketitle

\begin{abstract}
Consider the estimation of an unknown parameter vector in a linear measurement model. Centralized sensor selection consists in selecting a set of $k_{\text{s}}$ sensor measurements, from a total number of $m$ potential measurements. The performance of the corresponding selection is measured by the volume of an estimation error covariance matrix. In this work, we consider the problem of selecting these sensors in a distributed or decentralized fashion. In particular, we study the case of two leader nodes that perform naive decentralized selections. We demonstrate that this can degrade the performance severely. Therefore, two heuristics based on convex optimization methods are introduced, where we first allow one leader to make a selection, and then to share a modest amount of information about his selection with the remaining node. We will show that both heuristics clearly outperform the naive decentralized selection, and achieve a performance close to the centralized selection.
\end{abstract}

\section{Introduction}

Consider a linear model where a centralized collector estimates an $n$-dimensional parameter vector via an arrangement of $m$ sensors. The sensor readings are affected by measurement noise. The noise samples are assumed to be realizations of independent identically distributed Gaussian random variables. Now suppose the collector is allowed to use $ k_{\text{s}}$  active sensors only, where $n \leq k_{\text{s}} <m$. We call such a situation a centralized sensor selection problem. The performance of a particular selection can be assessed by the volume of the estimation error covariance matrix \cite[Sec.~\RM{2}.A]{JoshiBoyd2009}. Therefore, the objective of the sensor selection problem is to select $k_{\text{s}}$ sensors such that this volume is minimized. For this purpose, the centralized collector must know the complete measurement matrix, which is needed for calculating the error covariance.

\begin{figure}[htp]
\includegraphics[width=0.915\columnwidth]{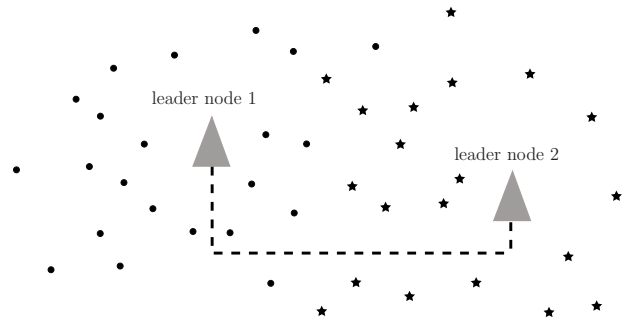}
\caption{Arrangement of different sensors and leader nodes. Sensors $(\bullet)$ are associated with leader node 1, sensors $(\star)$ with leader node 2. Both leader nodes may share a (very) limited amount of information.}
\label{fig:decentralized_model}
\end{figure}

In contrast, consider the sensor arrangement as depicted in Fig.~\ref{fig:decentralized_model}. We have a partition of all sensors into two groups. Each sensor group is associated with a specific leader node. The decentralized sensor selection problem consists in selecting a subset of sensors by the corresponding leader nodes individually. After that, the individual selections are transmitted to the centralized collector. The main advantage of such an approach is that we do not need to know the complete measurement matrix at one point, i.e., at the centralized collector. This can be motivated, for example, by limitations of the available transmission bandwidth in a sensor network. However, there is no guarantee that individual selections minimize the volume of the error covariance matrix. The reason is that the decentralized leader nodes may choose jointly correlated measurements, without even knowing it. In this work, we propose two simple heuristic methods for decentralized sensor selection. Both heuristics try to avoid jointly correlated measurements by transmitting a modest amount of data between leader nodes. We will show by numerous numerical experiments that the performance can be very close to the centralized solution.

The mathematical form of the sensor selection used in this paper was introduced in \cite{JoshiBoyd2009},\cite{YaoSetharesKammer1992}. In particular, the authors in \cite{JoshiBoyd2009} study the sensor selection problem embedded in the framework of convex optimization. Throughout this paper, we will make extensive use of this approach. In \cite{MoAmbrosinoSinopoli2009}, a multi-step sensor selection strategies based on the Kalman filter error covariance matrix is investigated. Other authors propose single sensor scheduling algorithms, e.g., \cite{Gupta2006},\cite{Vitus2010}. A different, but conceptually related approach is the selection of reliable sensors in the context of robust sensing \cite{KekatosGiannakis2010}. 

The remainder of the paper is organized as follows. Section \RM{2} introduces the centralized and decentralized sensor selection problem. To the best of our knowledge, the latter was not considered by others in this form. In Section \RM{3}, two heuristics for decentralized sensor selection are motivated and developed. The solution of the (nonconvex) sensor selection problem is outlined in Section \RM{4}. In Section \RM{5}, extensive numerical simulation illustrates the performance gains of the proposed decentralized heuristics.
 
\section{Sensor Selection}

\subsection{System Model}

A linear measurement model can be written as \cite[Sec.~\RM{2}.A]{JoshiBoyd2009}
\begin{align}
y_i & = a_i^T x + v_i, \qquad i = 1,\ldots,m
\end{align}
where $x\in\mathbb{R}^n$ is an unknown parameter vector that we want to estimate, $y\in\mathbb{R}^m$ is the measurement vector, and $m>n$. Throughout this paper, we will use the terms sensor and measurement synonymously. The measurements are corrupted by noise $v_1,\ldots,v_m$ that is independent and identically distributed (iid) with $\mathcal{N}(0,\sigma^2)$. The measurement matrix 
\begin{align}
A = \begin{bmatrix} a_1^T \\ \vdots \\ a_m^T \end{bmatrix}
\end{align}
is assumed to have full column rank, i.e., $\mathbf{rank}(A) = n$. The maximum-likelihood estimate of $x$ is then given by
\begin{align}
 x_{\text{ML}} & = \left(\sum_{i=1}^m a_i a_i^T \right)^{-1} \sum_{i=1}^m y_i a_i.
\end{align}
The covariance matrix of the estimation error $x-x_{\text{ML}}$ has the form
\begin{align}
\label{eq:error_covariance_matrix}
 \Sigma & = \sigma^2 \left( A^T A \right)^{-1} = \sigma^2 \left(\sum_{i=1}^m a_i a_i^T \right)^{-1}.
\end{align}
We measure the quality of the estimation by the volume of this matrix. It can be shown \cite[Sec.~\RM{2}.A]{JoshiBoyd2009} that this  measure is related to the log-volume of a confidence ellipsoid given by
\begin{align}
\log \mathbf{vol}(\mathcal{E}) & = \text{const.} - \frac{1}{2} \log \det \left(\sum_{i=1}^m a_i a_i^T \right).
\end{align}
This volume is a scalar measure for how informative the measurements are or how uncertain we have to be about our estimate $x_{\text{ML}}$. In particular, a small volume corresponds to a small uncertainty, and vice versa. 

\subsection{Centralized Sensor Selection Problem}
\label{sec:centralized_ss}

Now suppose we have a total number of $m$ measurements. A central collector attempts to find a subset of $k_{\text{s}} < m$ measurements that minimizes the uncertainty about $x_{\text{ML}}$. This leads to the \textit{centralized sensor selection problem} that is stated in \cite[Sec.~\RM{2}.B]{JoshiBoyd2009} as
\begin{equation}
\label{eq:centralized_ss}
\begin{array}{ll} 
\underset{z}{\text{maximize}} & f_{\text{cen}}(z) = \log \det \left( \displaystyle\sum_{i=1}^m z_i a_i a_i^T \right) \\
\text{subject to} & \vecone^T z = k_\text{s} \\ & z_i \in \{0,1\}, \hspace{0.5cm} i = 1,\ldots,m
\end{array}
\end{equation}
where $\vecone$ is a vector of appropriate dimension with all entries equal to one. Each Boolean variable $z_i$ corresponds to a particular choice of a measurement. Whenever $z_i=1$, the $i$th measurement is to be used. The linear constraint $\vecone^T z = k_\text{s}$ is a budget constraint on the total number of active sensors. On occasion, we will rewrite the objective as
\begin{align}
f_{\text{cen}}(z) & = \log \det \left(A^T \mathbf{diag}(z) A \right),
\end{align}
where the matrix $A^T\mathbf{diag}(z) A$ is assumed to be positive definite \cite[Ch. 7]{HornJohnson1990}.

Due to the Boolean constraints in \eqref{eq:centralized_ss}, the centralized sensor selection problem is a nonconvex optimization problem and is generally hard to solve. However, note that the objective is a concave function for $z_i \geq 0$ \cite[Sec. 3.1.5]{cvx_book}. Relaxing the Boolean constraints to $0\leq z_i \leq 1$, the \textit{relaxed centralized sensor selection problem} has the form
\begin{equation}
\label{eq:centralized_relaxed_ss}
\begin{array}{ll} 
\underset{z}{\text{maximize}} & f_{\text{cen}}(z) \\
\text{subject to} & \vecone^T z = k_\text{s} \\ & 0 \leq z_i \leq 1, \hspace{0.5cm} i = 1,\ldots,m
\end{array}
\end{equation}
This problem is a convex optimization problem and hence can be solved efficiently, where the solution is denoted as $z_{\text{cen}}^\star$. The relaxation gives us a global upper bound for the centralized sensor selection problem \eqref{eq:centralized_ss}. The bound is given by
\begin{align}
\label{eq:global_upper_bound}
 \qquad U_{\text{cen}} = f_{\text{cen}}(z_{\text{cen}}^\star),
\end{align}
and is subsequently used as a global performance measure for decentralized strategies.

In order to obtain a feasible solution to \eqref{eq:centralized_ss} we apply a simple rounding scheme as suggested in \cite{JoshiBoyd2009}. In this scheme, the elements of $z_{\text{cen}}^\star$ are rearranged in descending order. After that, the $k_\text{s}$ largest elements are set equal to $1$, and the remaining elements to $0$. This gives us a suboptimal solution $\hat{z}_{\text{cen}}$ to \eqref{eq:centralized_ss}, along with a lower bound $L_{\text{cen}}=f_{\text{cen}}(\hat{z}_{\text{cen}})$. This lower bound and the corresponding duality gap $\Delta_{\text{cen}} = U_{\text{cen}}-L_{\text{cen}}$ will be used later in order to compare centralized and decentralized methods. In particular, when this gap becomes sufficiently small, $\hat{z}_{\text{cen}}$ is nearly optimal for problem \eqref{eq:centralized_ss}. Note that it is possible to apply more sophisticated rounding schemes \cite[Sec.~\RM{3}.E]{JoshiBoyd2009}. Since we focus on the comparison of centralized and decentralized strategies, we will use only the prescribed simple rounding.

\subsection{Decentralized Sensor Selection Problem}
\label{sec:naive_decentralized_ss}

Selecting sensors in a centralized fashion, the full measurement matrix $A$ must be known at one point, e.g., the centralized collector. Now suppose we have two leader nodes that have access to half of the measurements $m/2 \geq n$ via the partition 
\begin{equation}
\label{eq:measurement_partition}
\begin{array}{lll}
 A = \begin{bmatrix} A_1 \\ A_2 \end{bmatrix}, & \text{leader node 1: } A_1, & \text{leader node 2: } A_2.
\end{array}
\end{equation}
where $A_1, A_2 \in \mathbb{R}^{m/2\times n}$, and $\mathbf{rank}(A_1) = \mathbf{rank}(A_2) = n$. In the decentralized sensor selection problem considered here, both leader nodes are only allowed to select $k_\text{s}/2$ sensors each. For each leader node $l$ with $l\in\{1,2\}$, we first solve the relaxed optimization problems
\begin{equation}
\label{eq:decentralized_relaxed_ss}
\begin{array}{ll} 
\underset{z_l}{\text{maximize}} & f_{l}(z_l) = \log \det \left( \displaystyle\sum_{i=1}^{m/2} z_{li} a_{li} a_{li}^T \right) \\
\text{subject to} & \vecone^T z_l = k_\text{s}/2 \\ & 0 \leq z_{li} \leq 1, \hspace{0.5cm} i = 1,\ldots,m/2
\end{array}
\end{equation}
where $z_l^\star$ is the optimal solution of \eqref{eq:decentralized_relaxed_ss}. We call this approach \textit{naive decentralized sensor selection} or simply decentralized sensor selection. After computing $z_l^\star$, both leader nodes apply the simple rounding scheme separately in order to obtain the selections $\hat{z}_l$. Finally, both nodes transmit their selections to the centralized collector. 

\subsection{Suboptimality and Performance}
\label{sec:suboptimality_and_performance}

The (global) performance of any decentralized method has to be judged at the centralized collector based on the full problem, i.e., the centralized objective $f_{\text{cen}}$. The global upper bound is given by the expression in \eqref{eq:global_upper_bound}. For calculating the lower bound, we first stack the solution vectors
\begin{equation}
\begin{array}{ll}
\hat{z}_{\text{dec}} = \begin{bmatrix} \hat{z}_1 \\ \hat{z}_2 \end{bmatrix}.
\end{array}
\end{equation}
and insert them into the centralized objective
\begin{align}
 L_{\text{dec}} = f_{\text{cen}}(\hat{z}_{\text{dec}}) & = \log \det \left( \begin{bmatrix}A_1 \\ A_2 \end{bmatrix}^T \begin{bmatrix} \hat{z}_1 & \\ & \hat{z}_2 \end{bmatrix}\begin{bmatrix}A_1 \\ A_2 \end{bmatrix}\right) \nonumber\\ 
& =  \log \det \left( A^T \mathbf{diag}(\hat{z}_{\text{dec}}) A \right).
\end{align}
Our measure for any decentralized strategy is then given by the suboptimality gap
\begin{align}
\label{eq:suboptimality_gap}
 \Delta_{\text{dec}} & = U_{\text{cen}} - L_{\text{dec}}.
\end{align}
Furthermore, the set of feasible solutions of \eqref{eq:decentralized_relaxed_ss} is a subset of \eqref{eq:centralized_relaxed_ss}. Therefore, we can also conclude that $f_{\text{cen}}(z_{\text{cen}}^\star) \geq f_{\text{cen}}(z_{\text{dec}}^\star)$.

Interestingly, it is not possible to make general statements about the lower bounds $L_{\text{cen}}$ and $L_{\text{dec}}$, respectively. The reason is that the prescribed simple rounding scheme produces \textit{one} suboptimal solution for the centralized sensor selection problem \eqref{eq:centralized_ss}. In principle, it is possible that the rounding from the decentralized leader nodes results in a different suboptimal solution, which in turn achieves a higher lower bound. However, as numerical evaluation suggests (see Sec.~\ref{sec:numerical_examples}) this effect does not occur very often.
\section{Approach}

In this section, we will introduce two methods for solving the centralized sensor selection problem \eqref{eq:centralized_ss} in a partially decentralized manner. Partially decentralized means that we are willing to transmit a negligible amount of data from one leader node to the other. In particular, we allow the transmission of $N\ll m$ vectors of dimension $\mathbb{R}^{n}$. Without loss of generality, we assume that leader node 1 shares some vectors with leader node 2. These methods can be seen as simple heuristics that attempt to improve the lower bound $L_{\text{dec}}$ and, accordingly, the suboptimality gap \eqref{eq:suboptimality_gap}.

\subsection{Main Idea}
The main idea behind both heuristics can be described as follows. Assume we have measurements that are approximately collinear, i.e., rows from the matrix $A$ are weakly correlated. In the case of two rows, it follows from the error covariance matrix \eqref{eq:error_covariance_matrix} that
\begin{align} 
a_j a_j^T + a_k a_k^T & \approx (1+\gamma)a_j a_j^T, \qquad \gamma \in \mathbb{R}.
\end{align}
In the above expression, we have the sum of two rank-1 matrices that can be approximately be rewritten as a scaled version of one rank-1 matrix. This means, we have (effectively) lost one rank. Now recall that each binary variable $\hat{z}_i$ corresponds to a specific selection out of all rows from $A$. Since $\sum_{i=1}^m \hat{z}_i a_i a_i^T$ is symmetric, it is orthogonally diagonalizable and we can rewrite the objective \eqref{eq:centralized_ss} as
 \begin{align*}
\log\det \left(\sum_{i=1}^m \hat{z}_i a_i a_i^T\right) & = \log\det \left(U \Lambda U^T \right) = \sum_{i=1}^n \log \lambda_i, \nonumber
\end{align*}
where $\lambda_i$ are the (positive) eigenvalues of $A^T \mathbf{diag}(\hat{z})A$. Selecting many weakly correlated measurements via $\hat{z}$ will reduce the effective rank of the above matrix. This increases the number of small eigenvalues, and therefore increases the total volume of the corresponding confidence ellipsoid.

Unless we are allowed to use $k=m$ measurements, the centralized sensor selection will avoid correlated measurements since they do not add significantly to the objective in \eqref{eq:centralized_ss}. This situation is different for the decentralized sensor selection problem. Both leader nodes maximize their own ellipsoid, using only the local data $A_1$ and $A_2$, respectively. However, it is possible and likely that their individual selections are jointly correlated, which in turn leads to a smaller global ellipsoid. Now suppose leader node 1 shares some information about the largest contribution to its own ellipsoid. Leader node 2 would avoid picking the same contribution. This is the main idea behind both heuristics.

To be more specific, suppose only leader node 1 has solved the relaxed decentralized sensor selection problem \eqref{eq:decentralized_relaxed_ss}. The main contribution to the volume of the local ellipsoid is given by the largest eigenvalues of the matrix $A_1^T\mathbf{diag}(\hat{z}_1)A_1$. Denote $\lambda_1, \ldots, \lambda_{N} $ the $N$ largest eigenvalues and $u_1,\ldots,u_N$ the associated eigenvectors with that matrix. We will now introduce two heuristics that capitalize the influence of these eigenvalues and -vectors.

\subsection{Focused Diversity Method}
\label{sec:fdm_ss}

Leader node 2 modifies its data matrix $A_2$ and selection vector $z_2$ such that
\begin{align}
 A_{\text{fdm}} & = \begin{bmatrix} A_2 \\ \lambda_1 u_1^T \\ \vdots \\ \lambda_N u_N^T\end{bmatrix}, & \tilde{z} = \begin{bmatrix} z_2 \\ \tilde{z}_{m/2+1} \\ \vdots \\ \tilde{z}_{m/2+N} \end{bmatrix}.
\end{align}
We can now rewrite objective of the decentralized sensor selection problem for leader node 2 as 
\begin{align} & \log \det \left( A_{\text{fdm}}^T \mathbf{diag}(\tilde{z}) A_{\text{fdm}}\right) \\
= &\log \det \left( \displaystyle\sum_{i=1}^{m/2} \tilde{z}_{i} a_{2i} a_{2i}^T + \sum_{i=1}^{N} \tilde{z}_{i+m/2}\lambda_i^2 u_i u_i^T \right).
\end{align}
Leader node 2 must avoid the same directions or contributions that were already made by leader node 1. One way to achieve this is to set $\tilde{z}_{m/2+1} = \ldots =\tilde{z}_{m/2+N} = 1$. Basically, leader node 1 has already made a decision for the second node. The objective for leader node 2 has then the form
\begin{align}
 f_{2,\text{fdm}}(z_2) = \log \det \left( \displaystyle\sum_{i=1}^{m/2} z_{2i} a_{2i} a_{2i}^T + \sum_{i=1}^{N} \lambda_i^2 u_i u_i^T \right).
\end{align}
Hence the remaining relaxed optimization problem to be solved is given by
\begin{equation}
\label{eq:fdm_relaxed_ss}
\begin{array}{ll} 
\underset{z_2}{\text{maximize}} & f_{2,\text{fdm}}(z_2) \\
\text{subject to} & \vecone^T z_2 = k_\text{s}/2 \\ & 0 \leq z_{2i} \leq 1, \hspace{0.5cm} i = 1,\ldots,m/2
\end{array}
\end{equation}
where the solution is denoted as $z_{2,\text{fdm}}^\star := z_2^\star$. We call this heuristic \textit{focused diversity method}. Note that the underlying maximization problem is still concave. After solving, leader node 2 performs the simple rounding scheme in order to obtain $\hat{z}_{2,\text{fdm}}$. The upper and lower bounds for the focused diversity method are then calculated based on the vectors
\begin{equation}
\begin{array}{ll}
z_{\text{fdm}}^\star = \begin{bmatrix} z_1^\star \\ z_{2,\text{fdm}}^\star \end{bmatrix}, & \hat{z}_{\text{fdm}} = \begin{bmatrix} \hat{z}_1 \\ \hat{z}_{2,\text{fdm}} \end{bmatrix}.
\end{array}
\end{equation}

\subsection{Linear Penalty Method}
\label{sec:lpm_ss}

Another way to force leader node 2 to avoid the main directions from leader node 1 is to introduce a linear penalty for choosing similar measurements. This can be accomplished by adding an additional term to the objective given in \eqref{eq:decentralized_relaxed_ss}. Consider the quantity
\begin{align}
\label{eq:reg_quantity}
\underbrace{\left|\frac{a_{2i}^T}{\lVert a_{2i} \rVert_2} u_j\right|}_{\text{similarity}} \cdot  \underbrace{\frac{\lambda_j}{\lVert a_{2i} \rVert_2}}_{\text{relevance}} & = \left| a_{2i}^T \lambda_j u_j \right| \cdot  \frac{1}{\lVert a_{2i} \rVert_2^2},
\end{align}
where $i = 1,\ldots,m/2$, and $j = 1,\ldots,N$. Consider the LHS first. Whenever a measurement $a_{2i}$ has a direction that is similar to a main direction from leader node 1, the absolute value of the normalized inner product between $a_{2i}$ and $u_j$ will be 'large'. In order to account for a possibly higher contribution from $A_2$, we put an additional weight on this similarity, which is called relevance. Note that we only need to calculate the product $\lambda_j u_j$. Therefore, the RHS of \eqref{eq:reg_quantity} does not violate the restriction that only $N$ vectors from leader node 1 can be shared.

Using the quantity \eqref{eq:reg_quantity} we can now construct a penalty term. Denote the cost vector
\begin{align}
 c_i & = \sum_{j=1}^N \left| a_{2i}^T \lambda_j u_j \right| \cdot  \frac{1}{\lVert a_{2i} \rVert_2^2} \geq 0, \qquad i = 1,\ldots,m/2.
\end{align}
We rewrite the objective for leader node 2 as
\begin{align}
 f_{2,\text{lpm}}(z_2) & = \log \det \left( \displaystyle\sum_{i=1}^{m/2} z_{2i} a_{2i} a_{2i}^T \right) - \sum_{i=1}^{m/2} c_i z_{2i} \\ & = f_2(z_2) - c^T z_2,
\end{align}
where $c^T = [c_1 \cdots c_{m/2}]$. The resulting heuristic is called \textit{linear penalty method}, associated with the (relaxed) concave optimization problem
\begin{equation}
\label{eq:lpm_relaxed_ss}
\begin{array}{ll} 
\underset{z_2}{\text{maximize}} & f_{2,\text{lpm}}(z_2) \\
\text{subject to} & \vecone^T z_2 = k_\text{s}/2 \\ & 0 \leq z_{2i} \leq 1. \hspace{0.5cm} i = 1,\ldots,m/2
\end{array}
\end{equation}
The solution is called $z_{2,\text{lpm}}^\star:=z_2^\star$, and the resulting vectors for calculating the upper and lower bound are given by
\begin{equation}
\begin{array}{ll}
z_{\text{lpm}}^\star = \begin{bmatrix} z_1^\star \\ z_{2,\text{lpm}}^\star \end{bmatrix}, & \hat{z}_{\text{lpm}} = \begin{bmatrix} \hat{z}_1 \\ \hat{z}_{2,\text{lpm}} \end{bmatrix}.
\end{array}
\end{equation}
\section{Solving a Sensor Selection Problem}

In order to solve a concave maximization problem similar to \eqref{eq:centralized_relaxed_ss}, several methods are at hand. One could resort to optimization software like CVX \cite{CVX}. In our case, we have implemented a logarithmic barrier method, see \cite[Sec. 11.3]{cvx_book}. A full implementation will be made available on our website. The reason for choosing a barrier method is that it can be implemented without great effort. However, other interior-point methods, which are used in practice more often (for example, primal-dual methods), may solve the above problems within fewer iterations and higher accuracy.

From \cite[Sec. \RM{3} D]{JoshiBoyd2009}, the approximate objective of the relaxed sensor selection is given by
\begin{align}
\psi_{\text{cen}}(z)&\!=\! \log \det \left( \sum_{i=1}^m z_i a_i a_i^T\right)\! +\! \kappa \sum_{i=1}^m\!\left(\log(z_i)\!+\! \log(1\!-\!z_i) \right) \\
& = \log \det \left( \sum_{i=1}^m z_i a_i a_i^T\right) + \phi_{\kappa}(z).
\end{align}
The authors also give explicit expressions for the gradient $\nabla \psi_{\text{cen}}(z)$ and Hessian $\nabla^2 \psi_{\text{cen}}(z)$. This is needed in order to calculate the Newton step in the inner iteration of the barrier method \cite{cvx_book}. In our case, leader node 2 has a modified approximate objective, depending on the method (e.g., focused diversity or linear penalty method). For completeness, the explicit expressions are given below.

For the focused diversity method \eqref{eq:fdm_relaxed_ss}, the approximate objective of leader node 2 has the form 
\begin{align}
 \psi_{\text{fdm}}(z_2)&\! =\! \log \det \left(\displaystyle \sum_{i=1}^{m/2} z_{2i} a_{2i} a_{2i}^T + \sum_{i=1}^{N} \lambda_i^2 u_i u_i^T \right) + \phi_{\kappa}(z_2),
\end{align}
where $\phi_{\kappa} : \mathbb{R}^{m/2} \rightarrow \mathbb{R}$. Therefore, the gradient is given by 
\begin{align}
\nabla \psi_{\text{fdm}}(z_2) & = \mathbf{diag}\left( A_2^T W_{\text{fdm}} A_2 \right) + \nabla \phi_{\kappa}(z_2), 
\end{align}
where 
\begin{align}
 W_{\text{fdm}} & = \left(\displaystyle\sum_{i=1}^{m/2} z_{2i} a_{2i} a_{2i}^T + \sum_{i=1}^{N} \lambda_i^2 u_i u_i^T \right)^{-1}.
\end{align}
The Hessian is given by
\begin{align}
 \nabla^2 \psi_{\text{fdm}}(z_2) &\! =\! -\left( A_2^T W_{\text{fdm}} A_2 \right)\odot\left( A_2^T W_{\text{fdm}} A_2 \right) + \nabla^2 \phi_{\kappa}(z_2),
\end{align}
where $\odot$ is the Hadamard product.

In the case of the linear penalty method \eqref{eq:lpm_relaxed_ss}, the approximate objective reads as
\begin{align}
\psi_{\text{lpm}}(z_2) & = \log \det \left( \displaystyle\sum_{i=1}^{m/2} z_{2i} a_{2i} a_{2i}^T \right) - c^T z_2 + \phi_{\kappa}(z_2).
\end{align}
Hence the gradient has the form
\begin{align}
 \nabla \psi_{\text{lpm}}(z_2) & = \mathbf{diag}\left( A_2^T W_{\text{lpm}} A_2 \right) - c + \nabla \phi_{\kappa}(z_2),
\end{align}
where 
\begin{align}
 W_{\text{lpm}} & = \left(\displaystyle\sum_{i=1}^{m/2} z_{2i} a_{2i} a_{2i}^T \right)^{-1}.
\end{align}
Since we have a linear penalty term, the Hessian remains unchanged when compared to \cite[Sec. \RM{3} D]{JoshiBoyd2009}, except the different dimension $m/2\times m/2$.

\section{Numerical Example}
\label{sec:numerical_examples}

We will now compare the naive decentralized sensor selection (Sec.~\ref{sec:naive_decentralized_ss}) with the proposed heuristics, e.g., focused diversity (Sec.~\ref{sec:fdm_ss}) and linear penalty method (Sec.~\ref{sec:lpm_ss}). Our benchmark for all comparisons is the solution from the centralized sensor selection problem (Sec.~\ref{sec:centralized_ss}). 

Throughout this section we investigate the following modeling setup. We use $m = 100$ measurements, $n = 40$ unknown parameters, and a total number of $k_{\text{s}} = 40,\ldots,60$ sensors. The measurement matrix is partitioned as given in \eqref{eq:measurement_partition}. The entries of the submatrices $A_1$ and $A_2$ are iid with $\mathcal{N}(0,1)$. In order to create weakly correlated measurements, we pick randomly two different rows from $A_1$ and $A_2$, say row $i$ from $A_1$ and row $j$ from $A_2$. Both rows are then modified as follows
\begin{align}
a_{1i}^T & = \sqrt{1-\sigma^2} \cdot b^T +  \sigma w_i^T, \\
a_{2j}^T & = \sqrt{1-\sigma^2} \cdot b^T +  \sigma w_j^T,
\end{align}
where $b_1,\ldots,b_n$, $w_{i1},\ldots,w_{in}$, and $w_{j1},\ldots,w_{jn}$ are iid with $\mathcal{N}(0,1)$. In this formulation, $\sigma$ represents of the strength of correlation. In our simulation we modified a total number of 30 rows and used $\sigma = 0.1$. 

\begin{figure}
\centering
\subfigure[lower bounds $L$]{
\label{fig:lower_bounds}
\includegraphics[width=0.9\columnwidth]{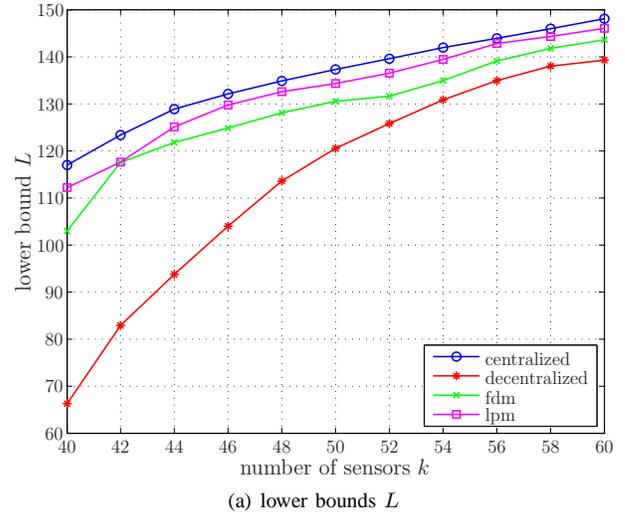}
} 
\subfigure[relative suboptimality gap $\Delta_{\text{rel},i}$]{
\label{fig:duality_gaps}
\includegraphics[width=0.9\columnwidth]{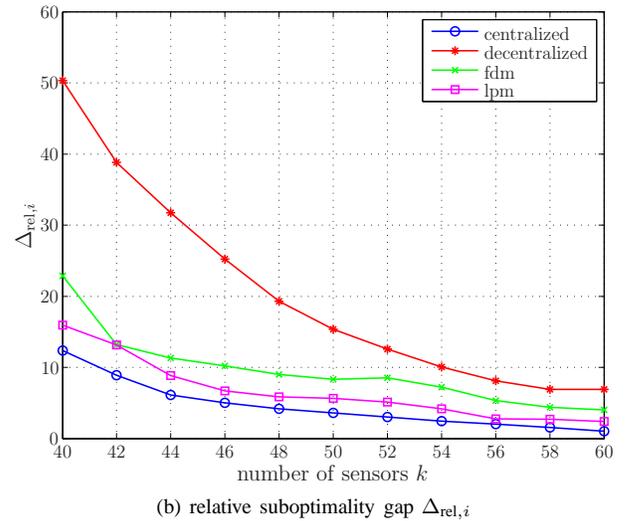}
} 
\caption{Lower bounds (a) and relative suboptimality gap (b) for the four different sensor selection strategies, where $N=5$. Note that the centralized sensor selection serves as a benchmark for all decentralized strategies.}
\end{figure}
We first fix the number of shared vectors $N=5$, and vary only the number of sensors to be used. In Fig.~\ref{fig:lower_bounds}, the lower bounds $L$ are shown for this case. We can clearly see that the focused diversity and the linear penalty method outperform the naive decentralized selection, notably for a low number of sensors. Since this gives only a lower bound we introduce the relative suboptimality gap
\begin{align}
 \Delta_{\text{rel},i} & = 100\cdot \frac{|U_{\text{cen}} - L_{i}|}{|U_{\text{cen}}|}, \qquad i \in \{\text{cen},\text{dec},\text{fdm},\text{lpm}\}.
\end{align}
This gap measures how far we are away from the optimum of the (nonconvex) centralized sensor selection problem \eqref{eq:centralized_ss}. Note that the relative suboptimality gap is a worst-case measure. This means in practice we are often closer to this optimum than suggested by this gap. The corresponding results are depicted in Fig.~\ref{fig:duality_gaps}. When compared to the naive decentralized selection, a considerable performance gain for both heuristics can be observed.

\begin{figure}
  \centering
  \subfigure[centralized]{
    \label{fig:gap_hist_cen}
    \includegraphics[width=0.84\columnwidth]{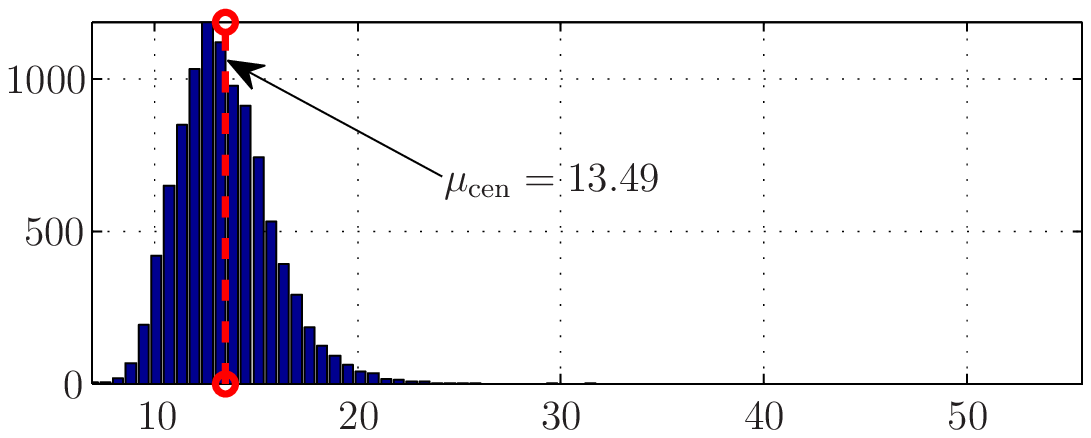}
  }
  \subfigure[decentralized]{
    \label{fig:gap_hist_dec}
    \includegraphics[width=0.84\columnwidth]{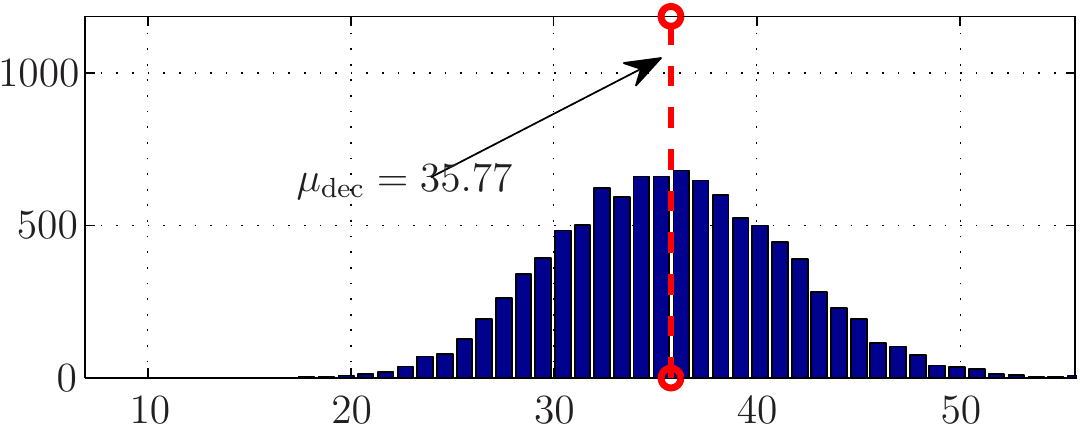}
  }
  \subfigure[focused diversity method]{
    \label{fig:gap_hist_fdm}
    \includegraphics[width=0.84\columnwidth]{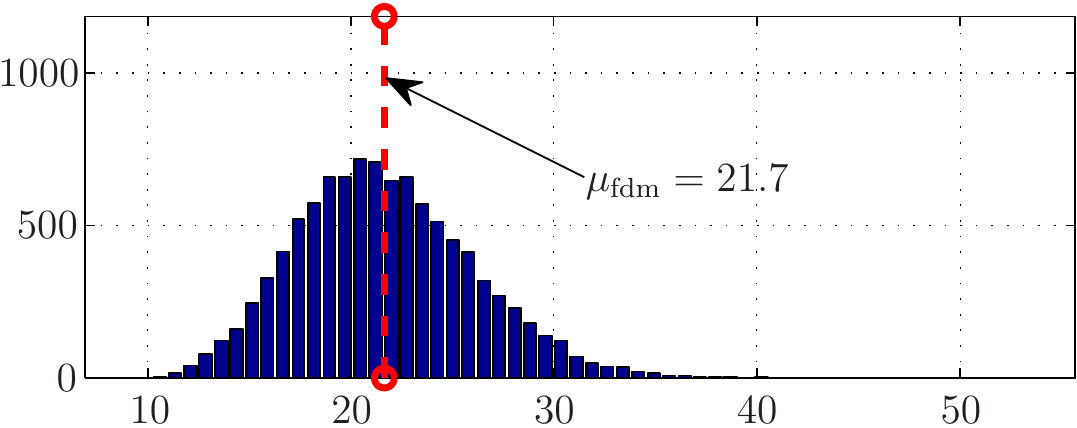}
  }  
\subfigure[linear penalty method]{
    \label{fig:gap_hist_lpm}
    \includegraphics[width=0.84\columnwidth]{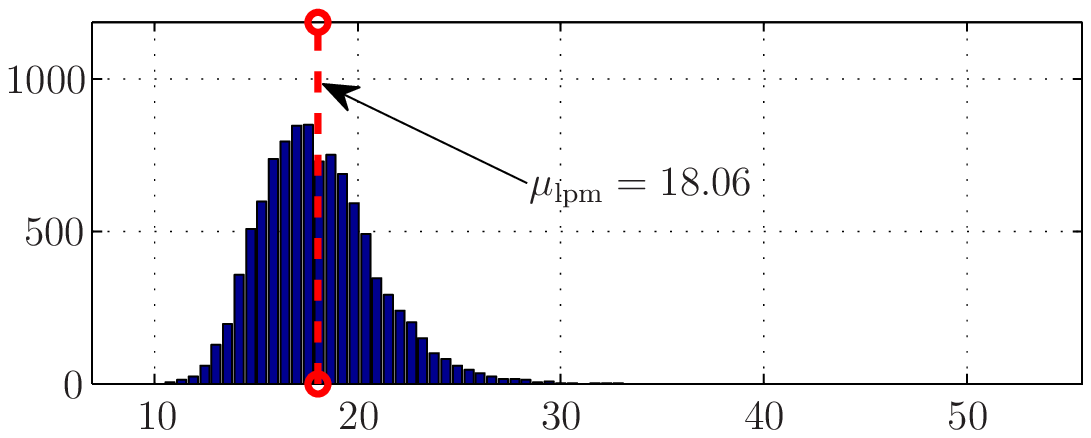}
  }
  \caption{Relative suboptimality gap for the four different sensor selection strategies, where $k_{\text{s}} = 40$ and $N=5$.}
  \label{fig:gap_hists}
\end{figure}

As mentioned in Sec.~\ref{sec:suboptimality_and_performance}, it is in principle not clear how the simple rounding scheme affects the different lower bounds, and hence the relative suboptimality gap. In order to get a meaningful interpretation, we run a simulation with $10^4$ different random realizations of the aforementioned modeling setup. The results for $k_s=40$ and $N=5$ can be seen in Fig.~\ref{fig:gap_hists}. As shown by the histograms, the linear penalty method performs on average slightly better than the focused diversity method. It is also evident that the decentralized sensor selection is far away from being optimal. We have also plotted the empirical means $\mu_i,$ of all relative suboptimality gaps.
\begin{figure}
\includegraphics[width=0.9\columnwidth]{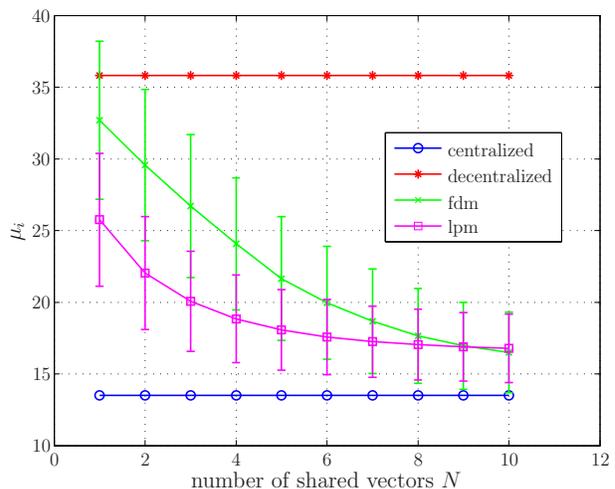}
\caption{Influence of the number of shared vectors $N$ on the empirical mean of the relative suboptimality gap, where $k_{\text{s}} = 40$. The error bars around the mean values represent the corresponding standard deviations.}
\label{fig:error_bars}
\end{figure}

The influence of the number of shared vectors on the relative suboptimality gap is illustrated in Fig.~\ref{fig:error_bars}. We assume that all vectors are transmitted perfectly, i.e., no transmission errors occur. Again, we have used total number of $10^4$ realizations and $k_{\text{s}} = 40$ sensors. Each point in Fig.~\ref{fig:error_bars} corresponds to a mean $\mu_i(N)$. As a reasonable deviation measure, we have included standard deviations depicted as error bars around the corresponding means. Note that we can not make the gap between the centralized selection and decentralized heuristics arbitrarily small. The reason is that leader node 1 already made a decision, which in turn can be suboptimal from a centralized point of view.

\section{Conclusion}

We have introduced the decentralized sensor selection problem for a partition of $m$ sensors between two leader nodes. A naive decentralized solution solves the sensor selection problem separately for both leader nodes. It was shown, that this solution is considerably worse when compared to a centralized approach. Therefore, we proposed two simple heuristics that allowed a very limited transmission of information from one leader to the other. These heuristics try to avoid similar contributions from both leader nodes, or equivalently, to exploit diversity in possible sensor selections. As suggested by extensive numerical simulations, the heuristics outperformed the naive decentralized selection by a substantial margin. 

\bibliographystyle{IEEEtran}
\normalsize
\bibliography{references}

\begin{thebibliography}{1}
\providecommand{\url}[1]{#1}
\csname url@samestyle\endcsname
\providecommand{\newblock}{\relax}
\providecommand{\bibinfo}[2]{#2}
\providecommand{\BIBentrySTDinterwordspacing}{\spaceskip=0pt\relax}
\providecommand{\BIBentryALTinterwordstretchfactor}{4}
\providecommand{\BIBentryALTinterwordspacing}{\spaceskip=\fontdimen2\font plus
\BIBentryALTinterwordstretchfactor\fontdimen3\font minus
  \fontdimen4\font\relax}
\providecommand{\BIBforeignlanguage}[2]{{%
\expandafter\ifx\csname l@#1\endcsname\relax
\typeout{** WARNING: IEEEtran.bst: No hyphenation pattern has been}%
\typeout{** loaded for the language `#1'. Using the pattern for}%
\typeout{** the default language instead.}%
\else
\language=\csname l@#1\endcsname
\fi
#2}}
\providecommand{\BIBdecl}{\relax}
\BIBdecl

\bibitem{JoshiBoyd2009}
S.~Joshi and S.~Boyd, ``Sensor selection via convex optimization,''
  \emph{Trans. Sig. Proc.}, vol.~57, pp. 451--462, February 2009.

\bibitem{YaoSetharesKammer1992}
L.~Yao, W.~Sethares, and D.~Kammer, ``Sensor placement for on-orbit modal
  identification of large space structure via a genetic algorithm,'' in
  \emph{Systems Engineering, 1992., IEEE International Conference on}, sep
  1992, pp. 332 --335.

\bibitem{MoAmbrosinoSinopoli2009}
Y.~Mo, R.~Ambrosino, and B.~Sinopoli, ``A convex optimization approach of
  multi-step sensor selection under correlated noise,'' in \emph{Communication,
  Control, and Computing, 2009. Allerton 2009. 47th Annual Allerton Conference
  on}, 30 2009-oct. 2 2009, pp. 186 --193.

\bibitem{Gupta2006}
V.~Gupta, T.~H. Chung, B.~Hassibi, and R.~M. Murray, ``On a stochastic sensor
  selection algorithm with applications in sensor scheduling and sensor
  coverage,'' \emph{Automatica}, vol.~42, no.~2, pp. 251 -- 260, 2006.

\bibitem{Vitus2010}
M.~Vitus, W.~Zhang, A.~Abate, J.~Hu, and C.~Tomlin, ``On efficient sensor
  scheduling for linear dynamical systems,'' in \emph{American Control
  Conference (ACC), 2010}, 30 2010-july 2 2010, pp. 4833 --4838.

\bibitem{KekatosGiannakis2010}
V.~Kekatos and G.~Giannakis, ``Selecting reliable sensors via convex
  optimization,'' in \emph{Signal Processing Advances in Wireless
  Communications (SPAWC), 2010 IEEE Eleventh International Workshop on}, june
  2010, pp. 1 --5.

\bibitem{HornJohnson1990}
R.~A. Horn and C.~R. Johnson, \emph{Matrix Analysis}.\hskip 1em plus 0.5em
  minus 0.4em\relax Cambridge University Press, 1990.

\bibitem{cvx_book}
S.~Boyd and L.~Vandenberghe, \emph{Convex Optimization}.\hskip 1em plus 0.5em
  minus 0.4em\relax New York, NY, USA: Cambridge University Press, 2004.

\bibitem{CVX}
M.~Grant and S.~Boyd, ``{CVX}: Matlab software for disciplined convex
  programming, version 1.22,'' \url{http://cvxr.com/cvx}, Feb. 2012.

\end{thebibliography}

\end{document}